\documentclass{desyproc}
\usepackage{slashed}

\begin{document}
\title{Renormalization of Fermion-Flavour Mixing}

\author{{\slshape Andrea A.~Almasy}\\[1ex]
Department of Mathematical Sciences, University of Liverpool, Liverpool L69 3BX, UK}

\contribID{xy}

\confID{800}  
\desyproc{DESY-PROC-2009-xx}
\acronym{LP09} 
\doi  

\maketitle

\begin{abstract}
We report on an explicit on-shell framework to renormalize the fermion-flavour mixing matrices in the Standard Model and its extensions, at one-loop level. It is based on a novel procedure to separate the external-leg mixing corrections into gauge-independent self-mass and gauge-dependent wave-function renormalization contributions. 
\end{abstract}

\section{Introduction}

Renormalizability endows the Standard Model (SM) with enhanced predictive power due to the fact that ultraviolet (UV) divergences from quantum effects can be eliminated by a redefinition of a finite number of independent parameters, such as masses and coupling constants. Furthermore, it has been known for a long time that, in the most frequently employed formulations in which the complete bare mass matrices of quarks are diagonalized, the Cabibbo-Kobayashi-Maskawa (CKM) quark mixing matrix must be also renormalized. In fact, this problem has been the object of several interesting studies over the last two decades. A matter of considerable interest is the generalization of these considerations to minimal renormalizable extensions of the SM.

\section{On-shell renormalization prescription\label{prescription}}

The on-shell renormalization framework we propose is a generalization of Feynman's approach in QED~\cite{Feynman:1949zx}. Recall that in QED the self-energy contribution to an outgoing fermion is given by
$$\Delta{\cal M}^{\rm leg} = \overline{u}(p)\Sigma(\slashed{p})\frac{1}{\slashed{p}-m}, \quad \Sigma(\slashed{p}) = A(p^2) + B(p^2)(\slashed{p}-m) + \Sigma^{\rm fin}(\slashed{p})$$
where $\Sigma(\slashed{p})$ is the self-energy, $A$ and $B$ are divergent constants, and $\Sigma^{\rm fin}$ is a finite part which is proportional to $(\slashed{p}-m)^2$ in the vicinity of $\slashed{p}=m$ and therefore does not contribute to $\Delta{\cal M}^{\rm leg}$. The contribution of $A$ to $\Delta{\cal M}^{\rm leg}$ is singular at $\slashed{p}=m$ and gauge independent and that of $B$ is regular but gauge dependent. They are called self-mass (sm) and wave-function renormalization (wfr) contributions. $A$ is cancelled by the mass counterterm $\delta m$ while $B$ is combined with proper vertex diagrams leading to a finite and gauge-independent physical amplitude. 

In the case of fermion-flavour mixing we encounter not only diagonal terms as in QED but also off-diagonal contributions. The self-energy corrections to an external fermion leg are now
$$\Delta{\cal M}_{ij}^{\rm leg}  = \overline{u}_i(p)\Sigma_{ij}(\slashed{p})\frac{1}{\slashed{p}-m_j},$$
where $i$ denotes the external fermion of momentum $p$ and mass $m_i$, and $j$ the virtual fermion of mass $m_j$. Using a simple algorithm that treats $i$ and $j$ on an equal footing, we write the self-energy as:
$$\Sigma_{ij}(\slashed{p}) = A_{ij}(p^2) + (\slashed{p}-m_i)B_{1,ij}(p^2)+B_{2,ij}(p^2)(\slashed{p}-m_j) + (\slashed{p}-m_i)\Sigma_{ij}^{\rm fin}(p^2)(\slashed{p}-m_j),$$
in analogy to QED. Similarly, we identify the contributions to $\Delta{\cal M}^{\rm leg}$ coming from $A$ as sm and those coming from $B_{1,2}$ as wfr contributions. Again, $\Sigma^{\rm fin}$ gives zero contribution. 

We consider next the cancellation of the sm contributions with the mass counterterms. We start from the bare mass term in the Lagrangian, $-\overline{\Psi}^\prime m^\prime_0\Psi^\prime$, and decompose the bare mass into a so-called renormalized mass and a corresponding counterterm, $m_0^\prime=m^\prime+\delta m^\prime$. We then apply a bi-unitary transformation on the fermion fields $\overline{\Psi}^\prime,\Psi^\prime$ that diagonalizes $m^\prime$ leading to the transformed mass term $-\overline{\Psi}(m+\delta m^{(-)}P_L+\delta m^{(+)}P_R)\Psi$. Here $P_{R,L}=(1\pm\gamma_5)/2$ are the chiral projectors, $m$ is real, diagonal and positive and $\delta m^{(\pm)}$ are arbitrary non-diagonal matrices subject to the Hermiticity constraint
\begin{equation}
\delta m^{(+)}=\delta m^{(-)\dagger}.\label{hermiticity}
\end{equation}
Further we adjust $\delta m^{(\pm)}$ to cancel, as much as possible, the sm contributions to $\Delta {\cal M}^{\rm leg}$.

The diagonalization of the complete mass matrix $M=m+\delta m^{(-)}P_L+\delta m^{(+)}P_R$ by means of a bi-unitary transformation of the form:
\begin{equation}
\psi_{L,R}=U_{L,R}\hat\psi_{L,R}\approx(1+ih_{L,R})\hat\psi_{L,R},\label{completediag}
\end{equation}
naturally induces a mixing counterterm matrix. Note that the second equality holds only at one-loop level. The matrices $h_{L,R}$ are chosen such that $\hat M$ is diagonal and are found to be:
\begin{equation}
i(h_{L,R})_{ij}=-\,\frac{m_i\delta m_{ij}^{(\mp)}+\delta m_{ij}^{(\pm)}m_j}{m_i^2-m_j^2},\qquad (h_{L,R})_{ii}=0.
\label{hmatrix}
\end{equation}

Due to the transformation in Eq.~(\ref{completediag}) the {$Vf_i\overline{f}_j$} bare interaction term in the Lagrangian transforms as well
$${\cal L}_{Vf_i\overline{f}_j}\propto\overline{\psi}_L^{f_i}K_0\gamma^\lambda\psi_L^{f_j}V_\lambda+H.c.\quad \xrightarrow{U_{L,R}}\quad  \overline{\hat\psi}_L^{f_i}(K+\delta K)\gamma^\lambda\hat\psi_L^{f_j}V_\lambda+H.c.,$$
with ${\delta K=i(Kh_L^{f_j}-h_L^{f_i}K)}$. $K_0=K+\delta K$ and $K$ are explicitly gauge independent and preserve the basic properties of the theory. $K$ is finite and identified with the renormalized mixing matrix. $\delta K$ is identified with the mixing counterterm matrix.

\section{Particular cases}

Following the procedure outlined in Sec.~\ref{prescription}, a CKM counterterm matrix was proposed in Ref.~\cite{Kniehl:2006bs}:
$$\delta V=i\left(Vh_L^D-h_L^UV\right),$$
with $h_L^{D,U}$ given by Eq.~(\ref{hmatrix}). Both $V_0=V+\delta V$ and $V$ satisfy the unitarity condition and are explicitly gauge independent. 

Some years later an alternative approach based on a gauge-independent quark mass counterterm expressed directly in terms of the Lorentz-invariant self-energy functions was proposed~\cite{Kniehl:2009kk}. 
The mass counterterms so defined obey three important properties: (i) they are gauge independent, (ii) they automatically satisfy the Hermiticity constraint of Eq.~(\ref{hermiticity}) and thus are flavour-democratic, and (iii) they are expressed in terms of the invariant self-energy functions and thus useful for practical applications.

A comparative analysis of the $W$-boson hadronic widths in various CKM renormalization schemes, including the ones discussed above, and the study of the implications of flavour-mixing renormalization on the determination of the CKM parameters are presented in Ref.~\cite{Almasy:2008ep}.

We have also considered the mixing of leptons in a minimal, renormalizable extension of the SM that can naturally accommodate heavy Majorana neutrinos. Here mixing appears both in charged- and neutral-current interactions and is described by the bare mixing matrices $B_0$ and $C_0$. Following ones more the prescription of Sec.~\ref{prescription}, we found that the charged-lepton mass counterterm is identical to that of quarks, up to particle content. However, in the case of the Majorana-neutrino there are two important modifications due to the Majorana condition $\displaystyle\nu=\nu^C$ (here $C$ denotes charge conjugation): (i) in addition to the Hermiticity constraint of Eq.~(\ref{hermiticity}) the mass counterterm should be symmetric, and (ii) now only one unitary transformation, $U^\nu=1+ih^\nu$, is needed to diagonalize the complete mass matrix $\hat M^\nu$. Keeping in mind the two changes, the mixing counterterm matrices are~\cite{Almasy:2009kn}
$$\delta B=i\left(Bh^\nu-h_L^lB\right)\qquad{\rm and}\qquad \delta C=i\left(Ch^\nu-h^\nu C\right).$$
Once $\delta B$ is fixed, $\delta C$ is fixed as well. Note that both, the bare and renormalized mixing matrices, are gauge independent and preserve the basic properties of the theory.

\section{Conclusions}

We proposed an explicit on-shell framework to renormalize the fermion-flavour mixing matrices in the SM and its extensions, at one-loop level. It is based on a novel procedure to separate the external-leg mixing corrections into gauge-independent sm and gauge-dependent wfr contributions. An important property is that this formulation complies with UV finiteness and gauge-parameter independence, and also preserves the basic structure of the theory.

\section*{Acknowledgements}

The author thanks B.A.~Kniehl and A.~Sirlin for the collaboration on the work presented here.

\begin{footnotesize}

\end{footnotesize}

\end{document}